\documentclass[aps,prm,twocolumn,superscriptaddress,floatfix,10pt]{revtex4}
\usepackage{graphicx}
\usepackage{amsmath}
\usepackage{xcolor}
\usepackage{wrapfig}
\usepackage{tikz}
\usepackage{latexsym}
\usepackage{appendix}
\usepackage{amssymb}
\newcommand{\tikzcircle}[2][black,fill=black]{\tikz[baseline=-0.5ex]\draw[#1,radius=#2] (0,0) circle ;}%

\begin{document}

\title{Tuning the thermal conductivity of silicon nanowires by surface passivation}

\author{C\'eline Ruscher}
%\email[]{cruscher@mail.ubc.ca}
\affiliation{Department of Mechanical Engineering, University of British Columbia, Vancouver BC V6T 1Z4, Canada}
\author{Robinson Cortes-Huerto}
\affiliation{Max-Planck Institut f\"ur Polymerforschung, Ackermannweg 10, 55128 Mainz, Germany}
\author{Robert Hannebauer}
\affiliation{Lumiense Photonics Inc., Vancouver, Canada}
\author{Debashish Mukherji}
\email[]{debashish.mukherji@ubc.ca}
\affiliation{Quantum Matter Institute, University of British Columbia, Vancouver BC V6T 1Z4, Canada}
\author{Alireza Nojeh}
\affiliation{Department of Electrical and Computer Engineering,University of British Columbia, Vancouver BC V6T 1Z4, Canada}
\affiliation{Quantum Matter Institute, University of British Columbia, Vancouver BC V6T 1Z4, Canada}
\author{A. Srikantha Phani}
\affiliation{Department of Mechanical Engineering, University of British Columbia, Vancouver BC V6T 1Z4, Canada}

\setlength{\parindent}{0pt}

\begin{abstract}
Using large scale molecular dynamics simulations, we study the thermal conductivity of bare and surface passivated silicon nanowires (SiNWs). 
For the cross--sectional widths $w \le 2$ nm, SiNWs become unstable because of the surface amorphosization and also due to the evaporation of a
certain fraction of Si atoms. {\color{black}The observed} surface (in--)stability is related to a large excess energy $\Delta$ of the surface Si atoms with 
respect to the bulk Si{\color{black}, resulting from the surface atoms being less coordinated and having dangling bonds.} 
{\color{black}We first propose} a practically relevant method that uses $\Delta$ as a guiding tool to passivate these
dangling bonds {\color{black}with hydrogen or oxygen, stabilizing the SiNWs. 
These passivated SiNWs are used to calculate the thermal conductivity coefficient $\kappa$.}
While the expected trend of $\kappa \propto w$ is observed for all SiNWs, 
surface passivation provides an added flexibility of tuning $\kappa$ with the surface coverage concentration $c$ of passivated atoms.
{\color{black}Indeed,} with respect to the bulk $\kappa$, passivation of SiNW reduces $\kappa$ by 75--80\% for $c \to 50\%$
and recovers again by 50\% for the fully passivated samples.
Analyzing the phonon band structures via spectral energy density, we discuss separate contributions from the surface 
and the core to $\kappa$. {\color{black}Our results also reveal that surface passivation increases SiNW stiffness, 
contributing to the tunability in $\kappa$.}
\end{abstract}

\maketitle

\section{Introduction}
\label{sec:intro}

Silicon (Si) is an important building block in designing nanoscale devices~\cite{PhysRevB.61.2651,Lieber02Nat,Dong08NL,Jo09NL,Mikolajick2013,ZhangRRL2013}, 
where tuning its physical properties, {\color{black}such as thermal and/or electrical transport, and mechanical stability,} 
is an essential for their use in a desired application. In this context, the ability to conduct the heat current 
is one of the key properties that dictates the applicability of a material under a wide range of environmental conditions~\cite{CahillRev}. 
For example, a high thermal transport coefficient $\kappa$ is needed for a heat sink material, while a low $\kappa$ is required 
for an high efficiency thermoelectric device \cite{Boukai2008,Jeong12JAP,Pennelli20NL,Li19CR}.

To achieve a desired (predictive) level of tunability in $\kappa$, silicon nanowire (SiNW) sensors are of
particular interest~\cite{Boukai2008,Mikolajick2013,ZhangRRL2013}. In particular, decrease in $\kappa$ was observed 
by changing the morphology via introducing kinks or pores \cite{JiangACSNanoLetters2013, Cartoixa2016, ZhaoKink2019}, roughness \cite{Hochbaum2008}, 
and the cross-sectional shapes \cite{LiuJAP2010}.
Tuning the mechanical property of SiNWs has also shown promising results, which may be achieved by introducing point defects in the crystalline lattice,
and thus $\kappa$ was shown to reduce by about $70\%$~\cite{MurphyACSNanoLetters2014}.

In the ongoing quest toward structural miniaturization, there is also a need to better understand these materials at 
the nanoscale, especially when the cross--section width $w \le 10$ nm. 
For nano--materials, lateral miniaturization often leads to a significant variation 
in $\kappa$ \cite{Li2003, Hochbaum2008, Lim2012, Boukai2008, VolzChen1999, Sellen10JAP, Rashid2018}. 
In particular, under strong confinement a delicate balance between the bulk as well as the surface phonon propagation 
controls $\kappa$ \cite{Ponomareva2007,Donadio2009, Zhou2017}. For example, it has been experimentally shown that $\kappa$ of SiNWs 
can be reduced by over an order of magnitude with respect to the bulk Si when $w \approx 22$ nm~\cite{Li2003}.
However, a significant challenge here is to attain a stable SiNW structure in the simulations for the smaller $w$ values \cite{Yu2016,Zhou2017}. 
When $w$ is small, i.e., only about a few tens of unit cells, surfaces of a free--standing SiNW become amorphous~\cite{surfaceAmor}, while the
core remains fairly crystalline. In some cases, even the surface Si atoms evaporate. This behavior 
is a result of the dangling bonds on the surfaces that originate because of the less coordinated surface Si atoms, 
and is not only known for SiNWs in the context of the electronic--band--structure~\cite{Term08APL}, but is also observed for the stability of metal nanoparticles \cite{Robin2013,Robin2015}. 

A common practical treatment to eliminate surface disorder is by the passivation of the surface dangling 
bonds~\cite{He2005,CuiLieber2003,YOUNGKIM2003269,Ashour_2013}.
In this context, while there are several experimental studies highlighting the importance of surface passivation, to the best of our knowledge, 
simulation studies on the exact role of surface passivation on SiNW stability and its connection to $\kappa$ are rather limited.
Computational studies, however, dealt with surface--nitrogenation~\cite{SiNH2011EPL}
and engineered surface--amorphousization~\cite{Donadio2009}. In both cases, a layer of hydrogen atoms was introduced on the surfaces,
without explicitly discussing the effects of hydrogen. Furthermore, we also note in passing that using N on Si is a rather nontrivial practical 
task because they usually induce large differential stresses and may make a surface unstable. Therefore, a detailed structure--property relationship is needed that may 
be used for the rational understanding and functional design of engineered advanced materials with stable surfaces.

Motivated by the above discussion, we study the $\kappa$ behavior in SiNWs with the goals to: 
(1) establish a structure--property relationship in SiNWs, 
(2) study the surface stability by using the excess energy $\Delta$ as a guiding tool and the effect of passivation on $\Delta$, 
(3) show how the surface passivation can lead to additional flexibility in tuning $\kappa$, and 
(4) discuss separate contributions of the surface and the core to the $\kappa$ behavior as revealed by the phonon band structures.
To achieve these goals, we employ a large scale atomistic molecular dynamics simulation protocol.

The remainder of the paper is organized as follows: in Section~\ref{sec:method} 
we sketch a detailed discussion of the simulation model and method.
Results are presented in Section~\ref{sec:res} and finally conclusions are drawn in Section~\ref{sec:conc}.

\section{Model and methods}
\label{sec:method}

\begin{figure}[ptb]
    \includegraphics[width=0.31\textwidth,angle=0]{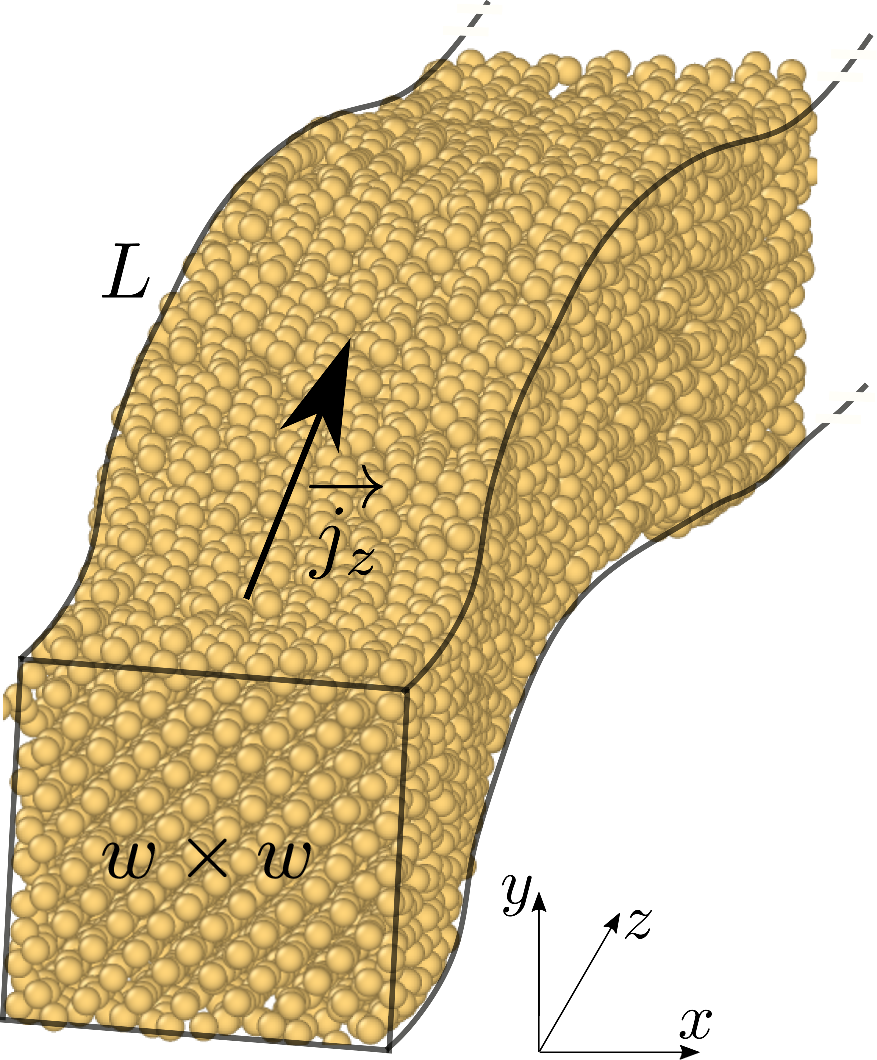}
    \caption{A simulation snapshot of an equilibrated bare silicon nano--wire (SiNW) 
    with a cross--section width $w = 4.34$ nm and a length $L = 69.52$ nm. 
    The thermal transport coefficient $\kappa$ is calculated along the $z-$direction 
    and $\overrightarrow{j}_{z}$ is the heat flux, which is the $\langle 1 \,0 \,0 \rangle$ crystalline plane oriented.}
    \label{fig:snapBare}
\end{figure}

Individual systems consist of free--standing SiNW of length $L = 128 a$ and the cross--section is chosen as a 
square geometry with the varying width $w$ that is taken between $4a$ and $24a$. Here, $a \simeq 0.54$ nm is the lattice constant.
This leads to the typical number of particles $N$ per system varying between $1.63 \times 10^4 - 5.89 \times 10^5$. 
We consider the $\langle 1~0~0 \rangle$ crystalline plane oriented along the $z-$axis, i.e., the direction of $\kappa$ calculation. 
A representative simulation snapshot of a bare SiNW highlighting the direction of heat flow is shown in Fig.~\ref{fig:snapBare}. The bare--SiNWs are also passivated with
oxygen (O--SiNW) or hydrogen (H--SiNW).

The initial configurations are created by placing a SiNW at the center of a simulation box with 
dimensions $50 w \times 50w \times L$. The periodic boundary conditions are applied along all directions, 
but a large vacuum in the lateral directions (i.e., along $x-$ and $y-$directions) 
prevents the Si atoms from seeing their  periodic images due to the flexural vibrations {\color{black}and thus precludes any simulation artifacts}. 

The Tersoff potential is used for the silicon (Si) interactions \cite{Tersoff1988}. A set of modified Tersoff 
parameters are used to mimic the interactions of Si with oxygen (O--SiNW) \cite{Munetoh2007} and hydrogen (H--SiNW) \cite{deBritoMota1999}.

The temperature $T$ is imposed using the Nos\'e-Hoover thermostat. 
Unless stated otherwise all physical properties are calculated at $T = 600$ K.
At different stages of simulations, spurious structural stresses along the $z-$direction are relaxed under NpT simulation, where pressure $p = 1$ atm is also imposed using the Nos\'e-Hoover barostat.
The equations of motion are integrated using the Verlet algorithm. The integration time step $\Delta t$ for SiNWs and O--SiNWs is $1.0$ fs,
while $\Delta t = 0.1$ fs is chosen for H--SiNWs. The GPU--accelerated LAMMPS package is used for these simulations \cite{Lammps2022}. 

\subsection{Equilibration of the free--standing SiNWs}

To achieve a stable free standing SiNW structure, the initial structural equilibration is performed in different steps:

\begin{enumerate}
    \item The crystalline SiNW structures are initially created at $T = 0$ K.

    \item Subsequently a SiNW is heated to $T = 10$ K for a time $t = 10^2$ ps. %under anisotropic NpT ensemble.

    \item The final configurations from step 2 are then heated further to $T = 600$ K for $t = 10^3$ ps. % in anisotropic NpT ensemble 
    For some test simulations, configurations are also created at $T = 300$ K.

    \item The configurations from the step 3 are then further equilibrated at $T = 600$ K for $t = 10^3$ ps. % and $p = 1$ bar under anisotropic $NpT$ 
    
\end{enumerate}

We note in passing that in steps 2--4, simulations are performed under an anisotropic pressure coupling that is only employed
along the $z-$direction, i.e., along the SiNW length. The final configurations obtained after step 4 are equilibrated 
in the canonical ensemble for an additional $t = 4 \times 10^3$ ps. We would also like to emphasize that a rather complex 
sample preparation protocol is used to ensure that the well--defined (residual stress--free) SiNW structures are obtained.

\subsection{Passivation of the free surface}

The surface passivated systems (O--SiNW and H--SiNW) are prepared at different surface coverage concentrations $c$. 
Here, $c = 1.0$ is defined as the maximum number of passivating atoms that can be added without introducing more dangling bonds by the
passivation. In the case of O--SiNW, one oxygen can bind to two Si atoms and thus leaves a small fraction of dangling bonds even for $c = 1.0$.
In H--SiNW, however, all dangling bonds are passivated because of their single Si--H coordination.
Subsequently, the samples with different $c$ are generated by randomly removing $(1 - c)$ fraction of passivating atoms.

\subsection{Thermal transport coefficient calculations}

The thermal transport coefficient $\kappa$ is calculated using the approach--to--equilibrium (ATE) method~\cite{Lampin2013}. Within this method, $L$ 
along the direction of heat flow is subdivided into three layers, i.e., $0-L/4$ (layer I), $L/4 - 3L/4$ (layer II) and $3L/4-L$ (layer III). 
In the first stage of canonical simulation, layers I and III are thermalized at $T_{\rm h} = T + 100$ K, while the layer II is kept at $T_{\rm c} = T - 100$ K.
After this thermalization stage, simulations are performed in the microcanonical ensemble that allows for the redistribution of energy, 
where $\Delta T = T_{\rm h} - T_{\rm c}$ are allowed to relax. As proposed in Ref.~\cite{Lampin2013}, we fit $\Delta T$ by a bi--exponential function
$\Delta T = c_1 \exp(-t/\tau_{1}) + c_{_{\parallel}}\exp(-t/\tau_{_{\parallel}})$ and obtain the time constant $\tau_{_{\parallel}}$ along the
direction of heat flow. Finally $\kappa$ can be calculated using,
\begin{align}
    \kappa = \frac{1}{4\pi^2} \frac{Lc_v}{w^2 \tau_{_{\parallel}}},
\end{align}
where $c_v$ is the heat capacity. We have used the Dulong--Petit classical estimate for $c_v = 3 N k_{\rm B}$, with 
$k_{\rm B}$ being the Boltzmann constant and $N$ is the total number of atoms in a system.

The calculations are performed at $T = 600$ K because the temperature equilibration in NVT 
and the subsequent relaxation in NVE for $\kappa$ calculations can be achieved within a reasonable 
simulation time without any unphysical energy drift. Note that the microcanonical simulations 
(especially with complex empirical potentials) typically show an energy drift over long times and for 
relatively large $\Delta t$~\cite{FSBook}. Furthermore, even when we are performing simulations at $T = 600$ K, 
the bare crystalline structure remains stable because $600~{\rm K} < \Theta_{\rm D}$ \cite{KittelBook}. 
Here, $\Theta_{\rm D}$ is the Debye temperature and for this model $\Theta_{\rm D} \geq 640$ K~\cite{Fan2020}.

The results of $\kappa$ for SiNW, O--SiNW, and H--SiNW are presented in comparison to the 
bulk thermal transport coefficient $\kappa_{\rm bulk}$. In our simulations, $\kappa_{\rm bulk}$ is calculated
in a $10a \times 10a \times 128a$ Si sample with PBC in all directions, i.e., without a large vacuum in 
the lateral directions. This calculated $\kappa_{\rm bulk}^{\rm Si}\simeq 10.1 \pm 1.3$ W/Km is consistent 
with an earlier simulation data in Ref.~\cite{Lampin2013} for $L \simeq 69.52$ nm, while is about a factor 
of six smaller that the known experimental value $\kappa_{\rm bulk}^{\rm exp}\simeq 65 $ W/Km \cite{Glassbrenner1964}. 
The observed variation between our simulation results and the experimental values is because $\kappa$ is $L-$dependent~\cite{Lampin2013,Bruns20prb} 
when calculated using a non--equilibrium method, as in the ATE method. This is particularly due to the reason that
the phonon wavelength is truncated along the direction of heat flow due to the boundary scattering. 
%This is a main reason why the observation of a length dependent $\kappa$ is expected. 
Furthermore, the central goal of our work is to investigate the change in $\kappa$ by passivation. 
Therefore, we do not go into more detail on the length effects on $\kappa$ and its value in the asymptotic limit. 
The latter can however be estimated to be $\kappa \simeq 145$ W/Km by extrapolating the data in Ref.~\cite{Lampin2013} 
for the same method and same model as ours. We have also attempted to calculate $\kappa_{\rm bulk}^{\rm Si}$ using 
the equilibrium Kubo--Green method~\cite{KuboG}, where the system size effects may be reduced. We, however, 
had severe problems with the convergence of heat flux auto--correlation function and drift in energy under NVE (data not shown).

\subsection{Spectral energy density calculation}

The phonon band structure is calculated using the spectral energy density (SED) $\phi({k}, \nu)$~\cite{Thomas2010,Thomas2015Erratum,Hornavar2016}. 
In our study, we define a supercell of Si atoms within a region of size $4a \times 4a \times a$ that in total has 
$N_{\rm supercell} = N_x N_y N_z = 128$ atoms per supercell. Here, $\phi({k}, \nu)$ is estimated using,
\begin{align}
\nonumber
    \phi(\bm{k}, \nu) = &\frac{1}{4 \pi \tau_0 N_T} \sum_{\alpha} \sum_{b}^{B} m_b \\
    &\left|\int_{0}^{\tau_0} \sum_{l}^{N_T} \dot{u}^{l,b}_{\alpha}(t) \exp(i \bm{k} \cdot \bm{r}_0^l -i\nu t) dt \right|^2, 
    \label{eq:SED}
\end{align}
where $\dot{u}^{l,b}_{\alpha}$ is the velocity of the $\alpha=\{x,y,z\}$ coordinate of the $b$ particle of mass $m_b$ in the unit supercell $l$. 
$r^{l}_0$ corresponds to the equilibrium position of the $l$ supercell. We note in passing that Eq.~\ref{eq:SED} is based on a normal mode 
analysis in the frequency domain that does not rely on the phonon eigenvectors, instead only on the 
Fourier transform of the atoms velocities~\cite{Feng2015} obtained over a total time $\tau_0$.
Using the equilibrated configurations, we run NVE simulations for $2$ns per sample with $\Delta t= 0.5$ fs for SiNWs and O--SiNWs and $\Delta t=0.05$ fs for H--SiNWs. 
Velocities and positions are stored every 16 fs.

\section{Results and discussions}
\label{sec:res}

\subsection{Surface stability and excess energy}

\begin{figure}[ptb]
    \centering
    \includegraphics[width=0.43\textwidth,angle=0]{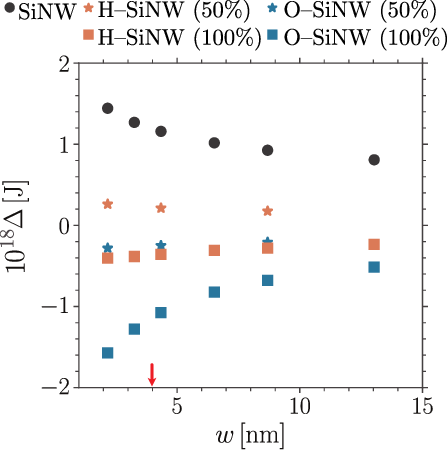}
    \caption{The excess energy $\Delta$ of the surface Si atoms as a function of the cross--section width $w$ of the silicon nanowires (SiNWs). 
    The data is shown for a temperature $T = 600$ K and for two different surface coverage concentrations $c$, 
    i.e., 50\% and 100\%. Red arrow indicates at a $w$ value that corresponds to the simulation snapshot in Figs.~\ref{fig:snapBare} and~\ref{fig:snapPassive}.}
    \label{fig:exeng}
\end{figure}

We begin by discussing the surface stability of SiNWs. For this purpose, we show a simulation snapshot of bare SiNW in Fig.~\ref{fig:snapBare}.
It can be appreciated that-- even when the core of SiNW is crystalline, there is visibly a large degree of
rearrangement of the surface Si atoms, forming an amorphous surface layer. 
To quantify this surface disorder and its correlation with the surface energy, we have calculated the excess energy using \cite{Cleveland1991,Baletto_etal_JCP116_3856_2002}
\begin{align}
\Delta  = \frac{E(N) - N\varepsilon^{\rm{Si}}_{\rm B}}{N^{2/3}}, 
\label{eq:excessenergy}
\end{align}
where $\Delta$ is calculated relative to the bulk cohesion energy per atom $\varepsilon^{\rm{Si}}_{\rm B}$ of bare Si and the energy of a SiNW with $N$ atoms is given by $E(N)$. 
The factor $N^{2/3}$ scales as the number of surface atoms. The variation of $\Delta$ with $w$ is shown in Fig.~\ref{fig:exeng}.

A closer look at the bare SiNW data (see the black \tikzcircle{2pt} data set) shows that $\Delta$ increases monotonically with decreasing $w$ 
and thus indicates that the surface effects become more dominant for smaller $w$. This $\Delta$ variation with $w$ is also directly 
related to the stability of SiNWs, especially for smaller $w$ where creating a surface usually has a significant energy cost.
Furthermore, we note that even when $\Delta$ decreases with $w$, it still remains significantly large for $w \to 13.0$ nm
(i.e., $\Delta \simeq 10^{-18}$ J or $\simeq 120~k_{\rm B}T$ at $T = 600$ K) and thus the surfaces remain fairly disordered 
for all $w$, see also the Supplementary Fig.~S1(a)~\cite{epaps}.

The central cause of such disorder is that the Si atoms are less coordinated on the surfaces than 
in the bulk and thus have a substantial number of dangling bonds. This contributes to a 
large energy penalty in forming a surface. Indeed, the surface energy, to a first approximation, is proportional to the total number of broken bonds at the surface~\cite{Robin2013}. 
Furthermore, it has also been shown that the Si--Si bond length decreases on the surface in comparison to the core \cite{Li_2014}. 
Such lattice contraction has been reported both in simulations and in experiments for Si nano-clusters \cite{Hofmeister1999,Yu2002}. 
Likewise, our simulations also reproduce this behavior of lattice contraction.
More specifically, cross--sectional surface area of a bare SiNW with $w = 4a$ is 
${\mathcal A}_{\rm cross} \simeq 4.05 \, \rm{nm}^2$, which is about $14 \%$ smaller than the expected value of $4.72 \, \rm{nm}^2$ for a perfect lattice. This has a direct implication on the elastic bending and torsional stiffness of the passivated SiNWs as we shall see later.

\begin{figure}[ptb]
   \includegraphics[width=0.49\textwidth,angle=0]{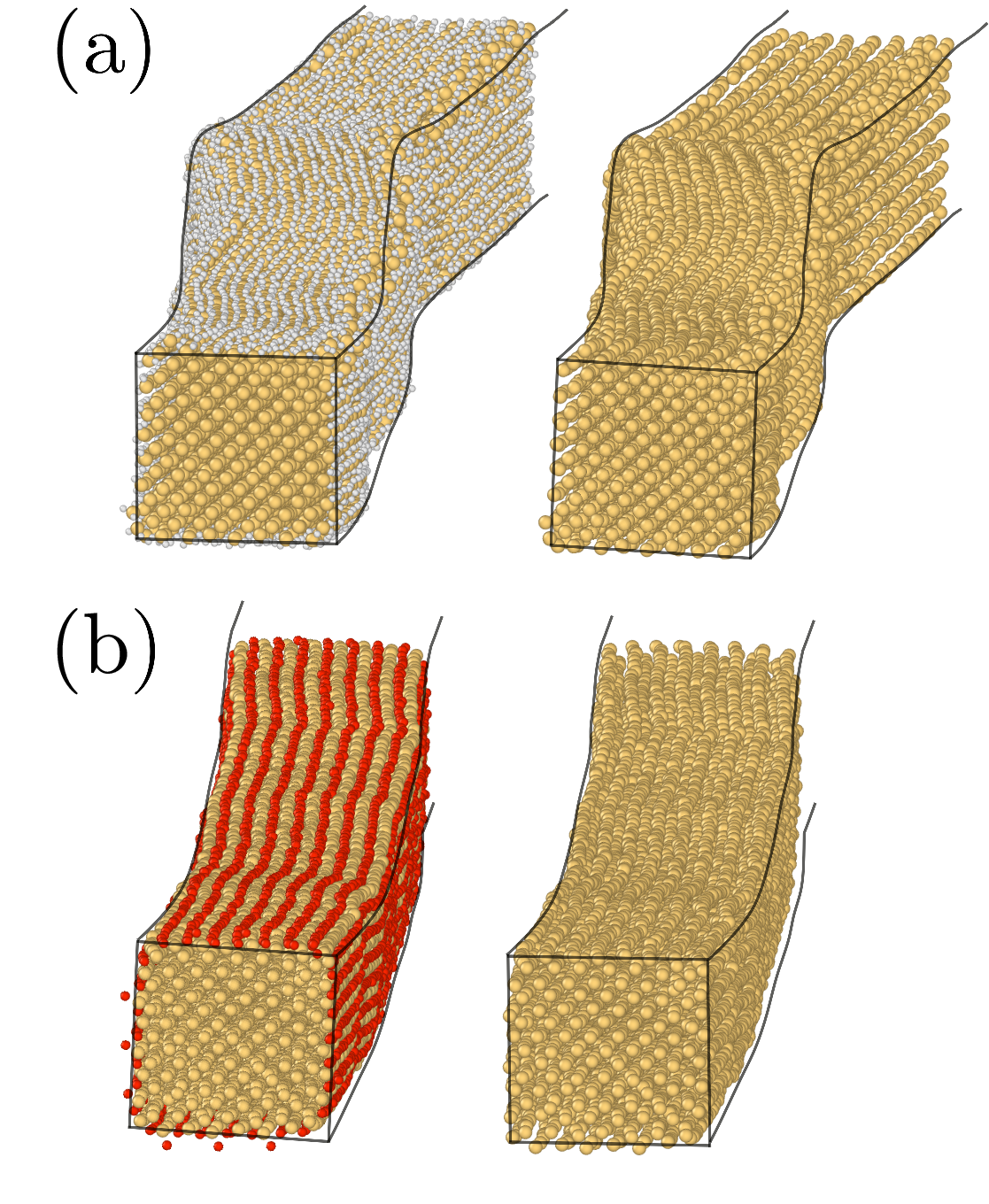}
    \caption{Simulation snapshots showing the passivated silicon nano--wire (SiNW) by hydrogen (part a) and oxygen (part b) for a cross--section width $w = 4.34$ nm. Snapshots are rendered with (left panels) and without (right panels) the passivated atoms. {\color{black}These simulation snapshots are shown for 100\% surface coverage concentration $c$ of the passivated atoms. 
    To obtain configurations at different $c$, corresponding numbers of passivated atoms are randomly removed from the surface.}}
    \label{fig:snapPassive}
\end{figure}

There are different treatments to eliminate surface disorder. Here, a common industrially relevant procedure is to deposit 5--10 nm of either silicon dioxide (SiO2) or silicon nitride (SiNH3) on bare SiNWs \cite{He2005, CuiLieber2003, YOUNGKIM2003269, Ashour_2013}. This serves as a natural route to reduce $\Delta$ as it reduces the number of dangling bonds. Simulations have also shown that passivation using amorphous Si \cite{Donadio2009} or hydrogen \cite{wolkowHpassivation,Li_2014} may  stabilize the surfaces. We take motivation from these experimental and simulation studies and passivate SiNW surfaces with either oxygen (O--SiNW) or hydrogen (H--SiNW). 
The latter is specifically chosen because hydrogen--on--silicon is used for inorganic solar cells\cite{Bertoni2011} and/or for data storage \cite{wolkowdata}. Note that we passivate the SiNW surfaces with only a monolayer of atoms. 

In Fig.~\ref{fig:snapPassive} we show 100\% passivated SiNWs, with (left panel) and without (right panel) 
rendering the passivated atoms. As expected, passivation helps preserving the crystallinity of 
the O--SiNW and H--SiNW surfaces (see the right panels in Fig.~\ref{fig:snapPassive}) in comparison to bare SiNW in Fig.~\ref{fig:snapBare}. 

Passivation also significantly reduces $\Delta$, see the $\star$ and $\blacksquare$ data sets in Fig.~\ref{fig:exeng}. It can be seen that $\Delta$ remains 
almost constant for all passivated systems, except for O--SiNW with 100\% passivation,
see blue $\blacksquare$ data set in Fig.~\ref{fig:exeng}. 
This observed large negative $\Delta$ for O--SiNW can be understood by assessing the different contributions in Eq.~\ref{eq:excessenergy}.
For example, $\Delta$ in Eq.~\ref{eq:excessenergy} is estimated relative to $\varepsilon^{\rm{Si}}_{\rm B}$, 
which only includes Si--Si binding energy. Here, the Si--Si binding energy is about 222 kJ/mol, while the Si--O binding energy can 
be as high as 452 kJ/mol \cite{HKK_InorganicChem}. These two separate contributions together will then result in a difference greater 
than just bare $\varepsilon^{\rm{Si}}_{B}$ in Eq.~\ref{eq:excessenergy}. This correction is then expected to lead to a smaller 
negative magnitude of $\Delta$ for O--SiNW. Note that these binding energy estimates are obtained at $T = 273$ K and our simulations are performed at $T = 600$ K.
Therefore, it may not be trivial to straightforwardly apply these estimates to our calculations.

\begin{figure}[ptb]
    \centering
    \includegraphics[width=0.4\textwidth]{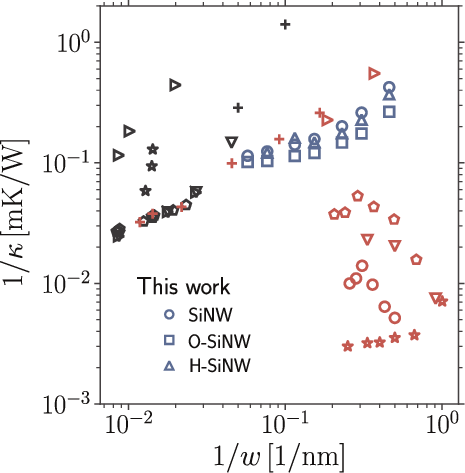}  
\caption{Inverse thermal transport coefficient $1/\kappa$ as a function inverse of cross--section width $1/w$ for different nanowires. 
The data from this work is shown for the bare silicon nanowires (SiNW), SiNW passivated with oxygen (O--SiNW), and SiNW passivated with hydrogen (H--SiNW), 
see the blue data sets. Results are only shown for the 100\% surface passivation coverage. 
$\kappa$ is calculated at a temperature of $T = 600$ K. For the comparison, we have also included data from the 
earlier published simulations (red data sets) \cite{VolzChen1999, Donadio2010, Ponomareva2007, Zhou2017, Lacroix2006, Rashid2018} and 
experiments (black data sets)\cite{Li2003, Hochbaum2008,Boukai2008,Lee2016,Lim2012}. We note in passing 
that we have also calculated the error bars from a set of six independent simulation runs for each configuration.
However, the error bars are not properly visible in a log--log plot with several decades on both the axes. 
Therefore, the raw $\kappa$ vs $w$ data (including their error bars) for all the samples are shown in the Supplementary Fig. S4.
    }
    \label{fig:kw}
\end{figure}

For H--SiNW, the reported H--H and Si--H binding energies are 432 kJ/mol and 318 kJ/mol, respectively~\cite{HKK_InorganicChem}. 
Other than explaining the negative $\Delta$ for smaller $w$, the difference might be at the origin of a competition between 
the passivation of the nanowire and the formation of a H$_2$ phase. We can expect that this competition becomes increasingly 
important as the temperature $T$ increases, possibly explaining the noticeable expansion of the surface area for H--SiNW. 
Interestingly, we note that despite the strong negative value of $\Delta$ for O--SiNW, which should promote the surface formation, 
the cross-section of the SiNW is actually shrinking, see the Supplementary Fig. S3~\cite{epaps}.
%This indicates that passivation affects the mechanical properties of nanowires in various way. 

\subsection{Thermal conductivity coefficient}

\begin{figure}[ptb]
    \centering
    \includegraphics[width=0.4\textwidth]{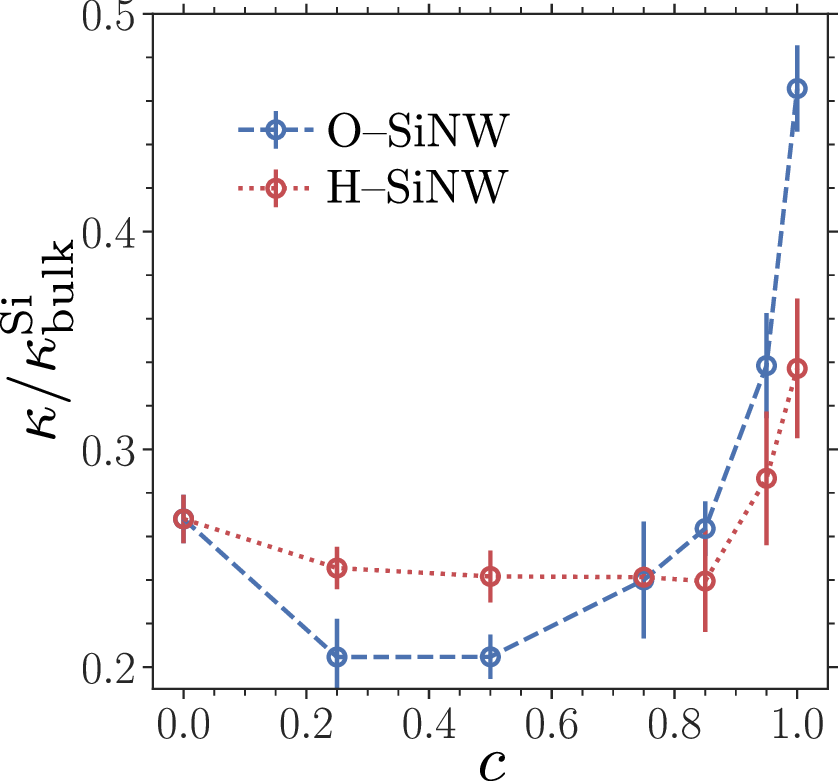}
    \caption{Normalized thermal transport coefficient $\kappa/\kappa^{\rm{Si}}_{\rm bulk}$ as a function of 
    surface coverage concentration of the passivated atoms $c$. Here $\kappa^{\rm{Si}}_{\rm bulk} = 10.1 \pm1.3$ W/mK. 
    Data is shown for a temperature of $T = 600$ K and for two different passivations, i.e., oxygen (O--SiNW) and hydrogen (H--SiNW).
    Individual $\kappa$ values are the averages of six independent runs 
and the error bars are standard deviation.
    }
    \label{fig:coverage}
\end{figure}

\begin{figure*}[ptb]
    \centering
       \includegraphics[scale=0.79]{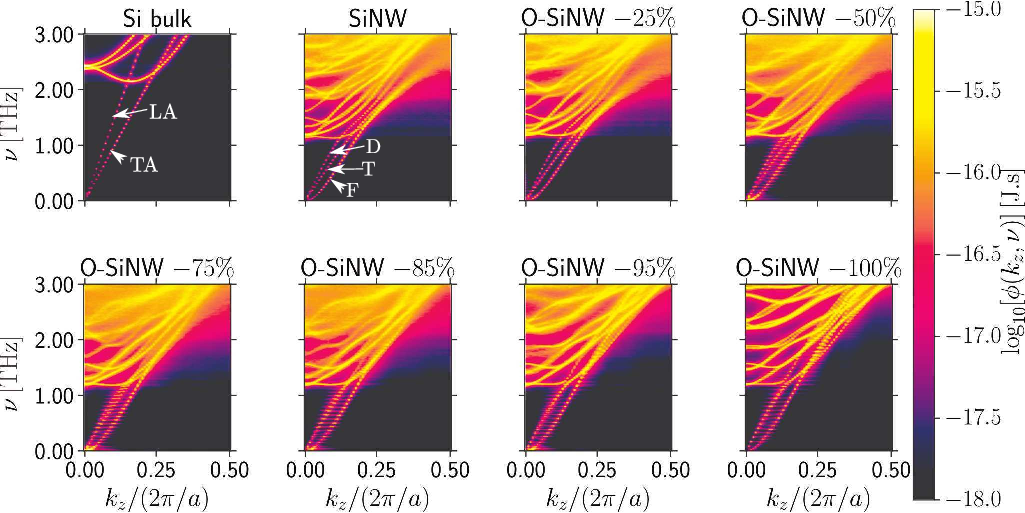}
            \caption{Spectral energy density (SED) of the oxygen passivated silicon nanowires (O--SiNW) for different surface coverage concentrations $c$. 
            For the reference, we have also included SED of bulk silicon and for a bare SiNW. Note that SED is calculated only including the silicon atoms.
            Data is shown for the smallest cross-section $w=4a$. For the clarity of presentation, we have annotated the longitudinal (L) and 
            transverse (T) acoustic branches for a bulk Si and the dilatational (D), the torsional (T), and the flexural (F) branches are highlighted for
            SiNW. Note that the D, T, and F branches can also be seen for the O--SiNWs. Moreover, the D and T branches merge for $c \ge 75\%$.
            }
    \label{fig:SED}
\end{figure*}

In Fig.~\ref{fig:kw} we show the variation of $\kappa$ with $w$. 
For comparison, we have also included earlier published simulation \cite{VolzChen1999, Donadio2010, Ponomareva2007, Zhou2017, Lacroix2006, Rashid2018} and experimental data \cite{Li2003, Hochbaum2008,Boukai2008,Lee2016,Lim2012}. 
Consistent with the experiments, our data also show the expected trend, i.e., $\kappa$ decreases monotonically with decreasing $w$. 
This is not surprising given that lateral miniaturization leads to an enhanced phonon scattering by the surfaces. 
It can also be seen that our simulation and some experimental data sets \cite{Li2003,Lee2016,Lim2012} show the same linear variation, i.e., $1/\kappa = 1/\kappa_{\infty} + {\alpha}/w$
with the same prefactor $\alpha$. This, however, we believe to be a mere coincidence because: (a) $L$ in experiments are significantly larger than in 
our simulations and $\kappa$ is known to increase with increasing $L$ \cite{Lampin2013,Bruns20prb}, especially for the quasi--one--dimensional materials. 
(b) A quantitative comparison between the computationally computed and experimentally observed $\kappa$ values is not possible
within the classical simulations.
The scenario (b) is due to the fact that in a classical simulation all modes contribute equally to $c_v$, while 
in reality many modes are quantum mechanically frozen below $\Theta_{\rm D}$ and thus do not contribute to $c_v$ \cite{binder99jpcb,Martin21prm}.
This leads to an overestimation of $c_v$ and thus also $\kappa$ in classical simulation in comparison to the experiments~\cite{KappaMDExp}.
Additionally, the phonon life time calculated within the classical simulations may also change due to the difference in the 
phonon distribution. It can also be seen that a few earlier simulations have shown opposite trends, i.e., $\kappa$ increases 
for $w$ below $\approx 2-3$ nm \cite{Ponomareva2007, Donadio2010, Zhou2017}. 
This non-monotonous change with decreasing $w$ was discussed in relation to the phonon-boundary scattering to hydrodynamic phonon flow,
which for the Tersoff potential, this transition has been located below 2 nm \cite{Zhou2017}. 
{\color{black}In this context, recent studies have shown that the hydrodynamic effects may play an important role~\cite{hydroKappa1,hydroKappa}. 
However, in our case, due to extreme confinement, these effects are negligible.} 

Note that the data sets presented in Fig.~\ref{fig:kw} are obtained at different $T$, $L$ and also by different methods. 
However, given that all all the data were obtained for $T < \Theta_{\rm D}$, the generic trend, i.e., $\kappa \propto w$, 
should remain the same, irrespectively of the system parameters and thus provides a robust overall picture of $\kappa$ in SiNWs.

Fig.~\ref{fig:kw} presents the importance of lateral miniaturization as a common protocol 
to tune $\kappa$ \cite{Li2003,Hochbaum2008,Lee2016,Boukai2008,Lim2012,Lacroix2006,Rashid2018}, 
while our results show that passivation provides an additional tuning parameter for $\kappa$. 
In Fig.~\ref{fig:coverage} it can be appreciated that $\kappa$ 
first decreases with increasing $c$ (for $c \leq 50\%$ for O--SiNW and $c \leq 75\%$ for H--SiNW), then again increases when 
$c \to 100\%$. It is also evident that O--SiNW shows a greater variation in $\kappa$ than H--SiNW.
What causes such non--monotonous variation in $\kappa$ with $c$? To answer this question, we focus on the phonon spectrum and 
compute the spectral energy density (SED) using~\cite{Thomas2010,Thomas2015Erratum,Hornavar2016}.

\subsection{Spectral energy density}

In Fig.~\ref{fig:SED} we show SED data for O--SiNW systems for different $c$. For reference, we have also calculated SED for bulk Si and bare SiNW. 
Fig.~\ref{fig:SED} shows several interesting features:

\begin{enumerate}

\item SiNWs have rather complex phonon band structures compared to a bulk Si. 
For $\nu < 1.0$ THz, SiNWs show the dilatational (D), the torsional (T), and the flexural (F) acoustic 
branches in the decreasing order of (slope) group velocity, respectively (see the labeling for SiNW). 
More importantly, the emergence of slow flexural phonons in the long wavelength 
acoustic limit can be seen, i.e., following $\nu \propto k_{z}^2$. 

\item The partial confinement of the optical phonons can also be observed, 
identified by the flat dispersion of the optical branches near $\nu \approx 1.2$ THz. Consistent with the literature, 
the D and T phonon branches (following $\nu \propto k_z$) become softer for SiNW than the bulk Si and thus also reduces $\kappa$~\cite{Cahill1992}.

\item The O--SiNW SED maps become sharper with increasing $c$, especially for $c \geq 75\%$. This is a clear indication 
of the improved crystallinity in O--SiNWs. The spectral width of the branches is also inversely proportional to the phonon lifetime $\tau_{\rm p}$~\cite{Thomas2010}, i.e.,
the sharper the SED map, the higher the $\tau_{\rm p}$ and thus an increase $\kappa$.

 \item Fig.~\ref{fig:SED} also reveal that the acoustic phonon branches steepen up with increasing $c$
and has a profound influence on the (intermediate) T branch., i.e., the T branch approaches 
the D branch with increasing $c$. Furthermore, there is a corresponding increase in the number 
of active phonon branches which changes the vibrational density of states $g(\nu)$ (see the Supplementary Figure S2)~\cite{epaps}. 

\item Passivation can also significantly change the flexural stiffness and hence group velocities (see the Supplementary Movie). 
The presence of the flexural stiffness significantly influence the phonon propagation. Larger flexural stiffness induces less number of kinks/bends along 
the O--SiNWs and thus have less number of phonon scattering sites for a given $L_z$~\cite{kinksSi2013NL,chen19jap,Mukherji21AN,mukherji24lang}. 

\end{enumerate}

To summarize the above observations-- when only a small amount of passivated atoms are deposited onto the surfaces
(i.e., $c \le 65\%$), they arrange randomly and improves crystallinity only in the local surface patches. This, however, 
does not introduce any long range order (as evident from rather diffused bands in Fig.~\ref{fig:SED} for $c \le 65\%$).
On the contrary, within this $c-$regime, the passivated atoms act as the surface defects for phonon scattering
and thus results in the initial decrease in $\kappa$ with $c$ in Fig.~\ref{fig:coverage}.
%which can not be compensated by the locally improved surface crystallinity. 

Further increase of $\kappa$ for $c > 65\%$ is because of two collective effects: 
(a) the improved overall crystallinity (as evident from Fig.~\ref{fig:SED}) and 
(b) stiffening of SiNWs via passivation, as indicated by the shift in the phonon peaks 
in $g(\nu)$ (see the Supplementary Fig. S2).

{\color{black}We note in passing that the observed non--monotonic trend in Fig.~\ref{fig:coverage} is a 
generic effect that is not only observed for a set of passivated SiNW, but rather any quasi--one--dimensional system
with the defect engineering via passivation shows the same trend. A typical example is the bottle--brush polymers, where
the concentration of side molecules controls the flexural stiffness and heat leakage along the backbone~\cite{mukherji24lang}.}

\begin{figure}
    \centering
       \includegraphics[scale=1.0]{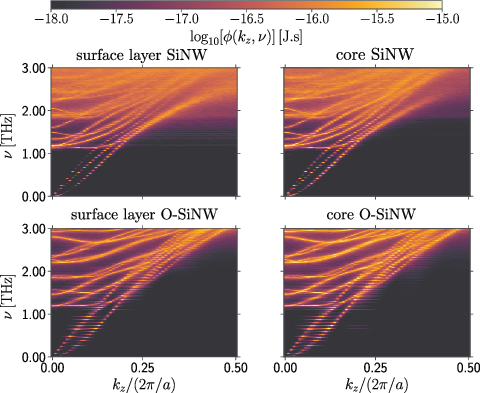}
            \caption{Spectral energy density $\phi(\bm{k}, \nu)$ of the surface (left panels) and core (right panels) Si atoms for SiNW and O--SiNW samples. 
            The data is shown for a cross--section width of $w = 4a$. The data is shown for $c = 100\%$ surface passivation.
            }
    \label{fig:SED_comparison}
\end{figure}

We would also like to highlight that $\kappa$ for H--SiNW is larger than for O--SiNW when $c < 80\%$ and show a shallower 
variation in $\kappa$ with $c$. While we can not provide an exact quantitative reason for this issue, we would like to point out that the H--SiNWs have slightly 
larger cross--section than O--SiNW, see the Supplementary Fig. S3. This indicates that the surface stresses are likely to be larger 
for H--SiNW samples and thus is consistent with the larger $\kappa$ values for $c < 80\%$. {\color{black}The increased $\kappa$
for O--SiNW for $c > 80\%$ is because of the increased flexural stiffness of a O--SiNW in comparison to a H--SiNW.}

We also comment that the surface passivation not only helps stabilize a SiNW, but it also 
improves the bulk crystallinity. To validate this we have calculated the SED contributions from the bulk and surface Si atoms separately. 
In Fig.~\ref{fig:SED_comparison} we show SED for a SiNW and a O--SiNW for $w = 4a$. It can be appreciated that the 
amorphous surface Si atoms also affect the bulk crystallinity (seen by the rather diffused SED branches in the top panels of Fig.~\ref{fig:SED_comparison}), 
while O--passivation improves surface crystallinity and thus also that of the bulk (see the bottom panels in Fig.~\ref{fig:SED_comparison}). 
Note that this behavior is most dominant for $w \leq 10a$. For the the larger $w$ amorphous Si surface does not significantly 
affect the bulk crystallinity, see the simulation snapshot in Supplementary Fig.~S1~\cite{epaps}.

\begin{figure}[ptb]
    \centering
      \includegraphics[scale=0.97]{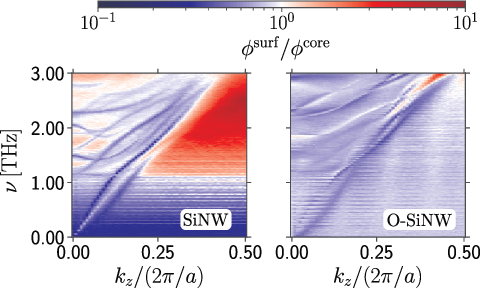}
    \caption{The ratio of spectral energy density (SED) computed for the Si atoms on the outermost layer of the nanowires with respect to the Si atoms in the core. Data is shown for the bare silicon nanowires SiNW (left panel) and oxygen-passivated nanowires O--SiNW (right panel). The data is shown for $c = 100\%$ surface passivation.}
    \label{fig:SEDratio}
\end{figure}

Lastly, to study the relative contribution of the surface modes to thermal transport, we have also computed $\phi^{\rm{surf}}/\phi^{\rm{core}}$. 
For the bare SiNWs (see the left panel in Fig.~\ref{fig:SEDratio}), it can be seen that D and F branches are dominated by the Si 
atoms at the core ($\phi^{\rm{surf}}/\phi^{\rm{core}} < 1$), while the surfaces and the core contribute equally to the T 
branch ($\phi^{\rm{surf}}/\phi^{\rm{core}} \simeq 1$). Note also that the red region within the 
range $k_z/(2\pi/a) \in [0.25;0.50]$ and $\nu > 1\, \rm{THz}$ is likely due to some background noise by the surface atoms.

The case of O--SiNW is somewhat different. For example, it can be seen from the right panel in Fig. \ref{fig:SEDratio} that the surface and the 
core contribute almost equally (i.e., $\phi^{\rm{surf}}/\phi^{\rm{core}} \simeq 1$). This explains that not only the surface crystallization that contributes to increase $\kappa$, %for 100\% passivated O--SiNW, 
instead a delicate balance between the surface crystallinity and the stiffness of an O--SiNW contributes to $\kappa$. 
A comparative comparison of the flexural vibrations between O--SiNW and H--SiNW is shown in the Supplementary Movie. 
This can be explained based on the fact that the passivation induces surface stress which increases the elastic torsional stiffness.

\section{Conclusion}
\label{sec:conc}

We have studied thermal transport $\kappa$ in surface passivated silicon nanowires (SiNW) using classical molecular dynamics. 
Consistent with experimental data, our results indicate that SiNWs become rather unstable and amorphous for small cross--section widths, 
which can be attributed to a large excess energy $\Delta$ of the surface silicon atoms. We used $\Delta$ as a guiding tool to stabilize 
the SiNW surfaces via surface passivation. As an added advantage, the concentration of passivation provides an additional tuning parameter 
for thermal conductivity. When the surfaces remain crystalline due to the passivated atoms, they also increase the stiffness of a SiNW and 
thus show that a delicate balance between different system parameters controls the tunability in $\kappa$. Using the phonon band structure analysis, 
we also decouple the surface and the bulk effects on $\kappa$.

The discussion presented in this work thus highlights that the $\kappa-$behavior of SiNWs and passivated SiNWs is dictated by a
delicate balance between different factors, i.e., surface--to--bulk crystallinity, flexural vibrations, and stiffness via passivation.
Therefore, it presents a microscopic picture of the heat flow in the surface treated low dimensional systems.\\

\noindent {\bf Supplementary Material:} This document contains supporting data for some of the claims made in the main manuscript text. In particular, simulation snapshots for the large cross--section SiNWs, vibrational density of states for SinW with passivation, and effect of passivation on SiNW cross--section are discussed.\\

\noindent {\bf Acknowledgement:} C.R. gratefully acknowledges MITACS and Lumiense Photonics Inc. for the financial support. 
C.R. further thanks Daniel Bruns for useful discussions and Pierre Chapuis for the help with the representations of SiNWs. 
We thank the generous allocation of GPU hours to D.M. at the ARC Sockeye facility where majority of simulations are performed.
Some simulations were also performed at the Compute Canada allocation to A.N.
For D.M. and A.N. this research was undertaken thanks, in part, to the Canada First Research Excellence Fund (CFREF), Quantum Materials and Future Technologies Program.\\

\noindent {\bf Data availability:} The scripts and the data associated with this research are available upon reasonable request 
from the corresponding author.\\

\noindent {\bf Author Contributions:} {C.R. wrote the LAMMPS scripts, ran the simulations, and analyzed the data. 
R.C.H. proposed the excess energy calculations and C.R. and R.C.H. co--analyzed the excess energy data. 
C.R. R.H., D.M., A.N., and S.P. proposed, designed, and conceptualized this study. 
C.R. and D.M. wrote the draft and all authors contributed to the editing. 
R.H., D.M., A.N., and S.P. contributed equally to this work and their names are written alphabetically.
D.M. wrote the reply to the referees and the revised draft.}\\

\noindent {\bf Competing interests:} The authors declare no competing interests.
%\bibliography{MukherjiJoP.bib}

\end{document}